# A Proposed Fuzzy Logic Approach for Conserving the Energy of Data Transmission in the Temperature Monitoring Systems of the Internet of Things

Noha Elqeblawy, Ammar Mohammed and Hesham A.Hefny

Department of Computer Science,
Faculty of Graduate Studies for Statistical Research, Cairo University, Egypt

## Abstract

*One of the primary challenges facing the Internet of Things is the reservation and efficient consumption of energy resources, especially in those types of applications that require continuous monitoring or suffer from lacking ongoing energy resources. Despite this, the indoor temperature and humidity monitoring systems are unconcerned about the insignificant amount of energy consumed during critical times when sending unimportant or useless data to the control room's servers. This paper proposes a fuzzy logic-based approach for reducing the amount of energy spent in indoor temperature and humidity monitoring systems by filtering data that is sent to servers based on several surrounding circumstances such as time of data recording and current energy consumption amount while maintaining constant monitoring. The experimental results on the Appliances Energy Prediction dataset show that the proposed fuzzy-based approach successfully reduces energy consumption in temperature and humidity monitoring systems by 11.8%.*

## 1. Introduction

The shortage of energy resources is turned to be a worldwide problem. Therefore, energy reservation and efficient consumption of energy is increasingly required in many applications and technologies [1]. The Internet of Things (IoT), which enables the distribution of smart objects all around us, is one of today's emerging technologies. These smart objects can communicate and share information with their corresponding servers via the internet to perform useful tasks without the need for human intervention [2]. The number of connected IoT devices is estimated to be more than 9 billion today, with a potential increase to 24 billion by 2024 [3].

Energy reservation could be viewed as a significant challenge for IoT, not only because of the global energy problem but also because of a lack of energy resources in several IoT applications [4], particularly those used in critical nonstop monitoring and control tasks or difficult-to-reach environments [5]. Such environments may suffer from a scarcity of ongoing energy resources capable of meeting application demands. Therefore, any interruption of service for recharging or changing its energy resources can obstruct their main function and leads to major problems.

For example, a hospital room for COVID-19 patients' needs to be remotely monitored all the time for safety aims while it is hard to be maintained frequently for infection prevention. A second example is when an IoT temperature monitoring system located inside a huge factory is interrupted due to energy resources running out, which may result in stopping the whole machine for maintenance works. As a result, it can have a significant impact on the manufacturing process. Furthermore, if that machine was intended to function for critical continuously running processes, such as electricity generation or water purification. As a result, any unplanned interruption to these machines can result in damage, service failure, significant financial loss, or even disasters. The third type of example is the natural disaster monitoring system that detects earthquakes or volcanic eruptions. If a disaster occurs, all normal recorded measured data before and during the disaster located in the surrounding area must be properly transmitted on time to take actions, such as alert, evacuation, or even search and rescue operations [6]. As a result, these critical data transmissions should not be hampered by a lack of energy resources in the applications, and energy consumption must be properly managed during the data transmission process. Data and control packets that periodical and frequent exchange between monitoring systems and their correspondent servers in IoT applications could be considered as one of the most energy-consuming tasks for these applications.

Frequently, the temperature and humidity monitoring systems contain sensors that detect the temperature and humidity degrees of the system's environment periodically. These sensors are linked to a microcontroller, which can interpret the recorded degrees locally and act accordingly, for example, by speeding up the local fan or turning off the air conditioner, as shown in [7]. Alternatively, they may transmit the measured temperature and humidity levels directly over the internet to their controlling server, where the data can be analyzed and used for prediction/forecast, research, statistics, or risk alarm purposes, as described in [8]. The main issue is the rapid and inefficient use of the system's energy resources that occurred during the data transmission process in temperature and humidity monitoring systems while transmitting each captured temperature or humidity degree at predefined time intervals directly to their controlling servers, regardless of whether those degrees are critical to be sent and stored in the current transmission environment. Furthermore, even if the uncritical degrees remain the same, they may transmit them multiple times during inconvenient transmission circumstances. Furthermore, the controlling server's storage resources will be depleted due to the frequent and redundant data received. This approach of data transmission could be referred to by the traditional approach.

To this end, the main contribution of this paper is to propose an approach that decreases energy consumption of data transmission in IoT temperature and humidity monitoring systems using fuzzy logic. Instead of transmitting all recorded degrees regularly, the fuzzy-based inference system is used to determine whether or not the captured temperature and humidity degrees should be transmitted to the server in the current transmission circumstances. The experimental results are carried out on the Appliances Energy Prediction dataset, which is publicly available on [9] to evaluate the performance of the proposed fuzzy-based approach.

Fuzzy logic is used to solve problems with imprecise facts that could be partially true and partially false at the same time that cannot be solved by classical logic as explained by [10,11]. In fuzzy logic, imprecise facts are represented by values in the interval [0,1]. The fuzzy set is defined by fuzzy logic as a set with no well-defined boundaries where the transition between membership and nonmembership is gradual rather than abrupt. A membership function is also used to define the fuzzy set. Various fuzzy sets are used to represent linguistic concepts/values, such as cold, cool, or warm. These linguistic concepts/linguistic values are used to define a variable's state. Such a variable is commonly referred to as a fuzzy variable/linguistic variable, for example, temperature, as referred to by [12,13],while the universe of discourse is defined as the domain of the fuzzy variable.

Generally, the architecture of fuzzy logic systems consists of four main parts, namely: fuzzifier, fuzzy rules, fuzzy inference engine, and defuzzifier. The fuzzifier is primarily in charge of the fuzzification process. Fuzzification is the process of using a membership function to convert input crisp values to fuzzy linguistic values with membership degrees. The fuzzification process's outputs, or fuzzified inputs, are then inferred by the fuzzy inference engine using a set of predefined fuzzy if-then rules to produce the output fuzzy variables. Finally, defuzzifier is in charge of the defuzzification procedure. The process of converting a fuzzy output variable to quantifiable crisp values is known as defuzzification. These precise values could be used to initiate events or actions following the system specifications outlined in [14,15].

This paper is organized as follows: Section 2 discusses related works found in the literature. Section 3 presents the proposed fuzzy logic monitoring systems, followed by section 4 that shows the experimental work. The results and discussions are in section 5, and finally, section 6 concludes the paper.

## 2. RELATED WORKS

Fuzzy logic has been used for energy-saving purposes in many IoT domains. For example, in the domain of the smart vehicle, Cueva-Fernandez et al. in [16] tried to save resources of the smart vehicle such as power and communication resources using fuzzy logic to choose the best moment for transmitting the data to the control room's server, depending on some parameters, such as vehicle position. Likewise, A Khalil et al. introduced in [17] an energy self-optimizing method for reducing energy utilization of IoT object data transmission. Based on Mamdani fuzzy logic system decisions, this method selects the best objects for transmitting the desired data while maximizing the corresponding IoT system lifetime. As inputs, the fuzzy logic system takes three IoT environment parameters (space between the IoT item gathering the data and the controller, remaining energy in the IoT item, and accuracy of the data collected by the IoT item). Furthermore, in the domain of healthcare monitoring, Y Mohammed et al in [18] introduced a fuzzy approach that assists in determining when and what patient data should be sent based on some of his basic health parameters from physiological IoT sensors. Their approach has the potential to reduce energy consumption by 29.35%. In the domain of smart homes, Giacomo Chiesa et al. presented in [19] an IoT system model with fuzzy logic to balance natural and artificial light in conjunction with a dynamic shading system for smart buildings to ensure comfort conditions and reduce artificial lighting energy desires.

Several research efforts tried to use fuzzy logic for saving energy in temperature monitoring systems that interpret the collected data locally as Diego et al. in [20] employed the IoT technologies and system multiagent to gather data of the environment of smart houses and decide on the best procedure in a central heating system of smart houses. The decisions are made by a smart agent that depends on fuzzy logic and the gathered data. They successfully reduce the energy consumption and financial costs. Also, Meana-Llorián, D., et al. in [21] introduced the IoFClime system, a temperature management system of a specific area that automatically gets the outdoor apparent temperature and humidity from external IoT platforms and senses the indoor temperature. Then, based on the collected data, they use fuzzy logic to determine the best time to heat up or cool down the atmosphere to maintain a comfortable temperature. As a result, this temperature management system saves approximately 40% of energy and thus saves money and electricity, which is similar to the work of Jana et al. in [22], who presented an indoor temperature controller using Gaussian type-2 fuzzy logic to save energy and create a more relaxing environment for users. Also, to save a significant amount of energy each year, DM Vistro et al. presented the IoT-ARTC Expert System in [23], which automates the temperature of the room using fuzzy logic. The system processes three input variables: temperature, pressure, and humidity to produce the output variables, heater, and chiller.

Similar to this work, there are a lot of works about temperature and humidity monitoring systems that transmit the sensed data to their control servers or external storage. Most of these works, however, are not concerned with energy conservation. Shabeeb et al. in [24] presented air temperature and humidity monitoring system for remotely monitoring the incubator for infants using an Arduino microcontroller, sensors, and open-source IoT applications. The system connects to a network via wireless fidelity (Wi-Fi) connection to connect to an application on a smartphone or computer. A portable physiological checking system, similar to the work of Valsalan et al. in [25], can frequently display the patient's health parameters and the temperature and humidity of the room. They proposed a continuous monitoring and control tool that would display the patient's condition and transmit the patient's information, as well as the temperature and humidity of the room, to the server.

On the other hand, Rizal et al. in [26] presented a real-time room circumstance remote monitoring system that measures room temperature and humidity, indicating that some researchers recognized the importance of energy conservation. The collected data is uploaded to an open-source platform where it can be retrieved and displayed via smartphones and web applications. However, their work used low hardware parts that may not be available for everyone.Furthermore, the system does not monitor all the time but it depends on the air-conditioner working time that may affect the room's safety, similar to the study of Shushan Qiao et al. in [27] who presented a temperature and humidity detection module system that concentrates on the low-energy consumption and high performance. The system is made up of two modules that transmit and receive temperature and humidity data.

Firmansah et al. in [28] presented their work on increasing the battery life of WSN in IoT indoor temperature and humidity monitoring systems using solar energy-harvesting technology. The system saves power by sleeping with periodic active breaks. Using solar energy is not always possible, especially during the winter season, and their results show that the daily gathered energy is still insufficient to meet the system's energy requirements. As demonstrated in previous works, most temperature and humidity monitoring systems that transmit data are unconcerned about efficient energy consumption or rely on hardware features to reduce energy consumption, in contrast to the proposed fuzzy-based approach, which attempted to avoid these issues by relying on ease of application and customizing the software solution to any temperature and humidity monitoring system requirements. Moreover, it is possible to apply for any existing system without the need for huge changes.

## 3. THE PROPOSED FUZZY LOGIC MONITORING SYSTEMS

Essentially, the proposed fuzzy logic-based approach aims to reduce the consumption of transmission energy in IoT temperature and humidity monitoring systems that send their collected data to the external storage. The key concept of energy reduction is to use fuzzy logic to reduce the energy consumed by packet exchange between temperature and humidity monitoring systems and server control rooms. Based on the surrounding environmental circumstances, fuzzy logic is used as a controller to determine whether or not the captured degrees should be transmitted to the server. This helps to avoid sending a large amount of useless data to the control server, which is especially important during critical times.

It should be noted that the required fuzzy logic system has to monitor four inputs, i.e., indoor temperature, indoor humidity, appliance consumption energy, and time of reading. The fuzzy system's (FS) output should be a decision to send or not send the data to the server to make the necessary adjustments. As shown in Figure 1, the system would function as a single-stage FS with five system linguistic variables, i.e., four inputs and a single output. Knowing that each linguistic variable should have a set of subjectively proposed linguistic values, the resulting fuzzy logic system will contain an excessive number of fuzzy if-then rules. For example, if each input has 4 fuzzy linguistic values, each one of them should appear with the other inputs' fuzzy linguistic values to produce all possible combinations of fuzzy rules. This results in a fuzzy rule base containing $4^4 = 256$ rules which considerably increases the complexity of the fuzzy system. In addition to this, the number of fuzzy rules will be greatly increased if there is a need to increase the number of inputs to the system. To avoid this situation, we propose decomposing the overall FS into three fuzzy subsystems organized in two stages corresponding to the input and output phases (see Figure 2). This approach allows the appropriate size of fuzzy if-then rules for each fuzzy subsystem and consequently the whole fuzzy model.

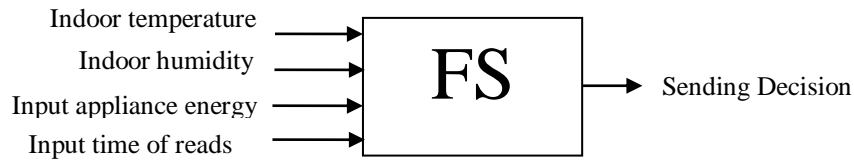

Figure 1. The overall fuzzy model without decomposition

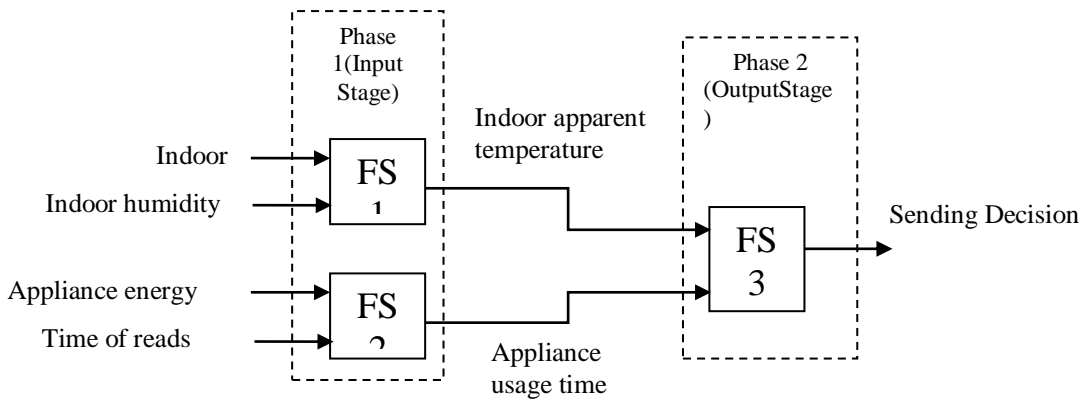

Figure 2. The proposed fuzzy model after decomposition

Therefore, the system configuration is shown in Figure 2 is adopted for building the fuzzy model to monitor the indoor area's temperature and humidity, which affect the overall apparent temperature. Higher degrees of apparent temperature recorded are given more transmission importance than lower degrees to maintain the safety of the indoor monitored area. The Mamdani fuzzy inference system is used in our model to clarify expertise in a more natural and human-like manner. The fact that the Mamdani model has fuzzy output compared with other fuzzy inference systems, makes it quite expressive and interpretable compared with other models that have crisp output rather than fuzzy output such as Sugeno-type FS [29]. The proposed fuzzy model is divided into two phases that determine whether or not the indoor temperature and humidity readings should be sent to the control server room. Phase 1 serves as an input fuzzy stage and is made up of two fuzzy subsystems, FS1 and FS2. The first input fuzzy subsystem is represented by FS1. It functions as a bank of fuzzy if-then rules that relate three parameters: two input linguistic variables: indoor temperature and indoor humidity, and one output linguistic variable: apparent temperature. The second input fuzzy subsystem, on the other hand, is represented by FS2. It functions as a bank of fuzzy if-then rules that also connect three parameters: two input linguistic variables, appliance energy read and reading time, and one output linguistic variable: appliance time. Phase 2, the output fuzzy stage, is made up of one fuzzy subsystem FS3. FS3 acts as a bank of fuzzy if-then rules that relate three parameters: two inputs which are the outputs of FS1 and FS2 and one output, which is the final decision of the whole fuzzy logic model. Choosing the four inputs: temperature, humidity, appliance energy, time of reading is mainly since they are the main features that affect the decision of sending as follows:

- Temperature and humidity determine the indoor apparent temperature that refers to increasing the importance of sending as long as the indoor apparent temperature is increased.
- Appliance energy refers to the amount of currently consumed energy which is inversely proportional to the ability to send. This means that the more amount of energy is used at this time, the more critical level of energy consumption and the less ability to send at this time. The higher apparent temperature is only allowed to be sent.
- The time of reading is an indication of the time during the day. For example, 2 pm is a peak period during the day in which everybody is awake at home or work which refers to high energy consumption at this time. High energy consumption critically reduces the ability to be sent, So, at this time, the higher apparent temperature can only be sent.

Figure 2 shows the structure of the proposed fuzzy model. Table 1 summarizes all the input and output linguistic variables of the fuzzy monitoring model.

Table 1. The input/output parameters of the proposed fuzzy model.

| System Parameters (Linguistic Variables) | Type | Source |
|---|---|---|
| Indoor Temperature | Input (phase1-FS1) | Internal temperature sensor |
| Indoor Humidity | Input (phase1-FS1) | Internal humidity sensor |
| Appliance energy | Input (phase1-FS2) | m-bus energy meters |
| Time of reads | Input (phase1-FS2) | Time of reads |
| Indoor apparent temperature | Input (phase2-FS3) | Output of phase1-FS1 |
| Appliance usage time | Input (phase2-FS3) | Output of phase1-FS2 |
| Sending decision | Output | Output of phase2-FS |

## 4. EXPERIMENTAL WORK

The proposed fuzzy monitoring model is implemented and tested using a benchmark dataset. As stated in the related works section in [18,21], the evaluation and energy reduction percentage isthe results of a comparison of traditional/conventional approaches to the proposed approaches. The same comparison technique is used in this paper. As a result, the model is evaluated by comparing it to the traditional approach for monitoring IoT temperature and humidity.

## 4.1. The Benchmark Dataset

The Appliances Energy Prediction dataset is a real-world dataset that has been presented by the authors of [9] by deploying IoT sensors inside a house located in Stambruges, Belgium. The dataset contains measures of temperature and humidity of various rooms (e.g., kitchen, laundry, bathroom, and ironing room). The Zigbee wireless sensor network recorded such measurements. Each wireless node sent temperature and humidity readings every 3.3 min, then store averaged every 10 min. A nearby weather station provided the weather data. The energy consumption data, on the other hand, were logged every 10 min via m-bus matter, as well as the energy used to light the house's fixtures. The data migration is based on the date and time, and the dataset has a period of 137 days (4.5 months). The dataset includes 29 distinct features and 19,735 observations/records, as shown in [9].

## 4.2. Calculation of Energy Consumption

For calculating the amount of energy that is supposed to be consumed during the transmission data packets process, equations (1) and (2) are adopted from [30]. Equation (1) calculates E(p) as the energy required to transmit a packet p.

$$E(p) = i \times v \times t_p \qquad (1)$$

Where: i is the current consumption, which is equal to 280 mA in transmit mode, and v is the used voltage, which is equal to 5 volts, assuming the usage of the IEEE 802.11g network interface card, as highlighted by [31].

Then, $t_p$ is the required time in seconds to transmit a packet given by the equation (2).

$$t_p = \left( \frac{p_h}{6 \times 10^6} + \frac{p_d}{54 \times 10^6} \right) \qquad (2)$$

Where: $p_h$ is the packet header size in bits, and $p_d$ is the packet data size in bits.

## 4.3. The Fuzzy Model Implementation

The proposed fuzzy model, shown in Figure 2, is implemented using MATLAB Toolbox [32]. Figure 3 shows the implemented system configuration. We proposed a group of fuzzy sets with specific forms of membership functions as linguistic values to each fuzzy linguistic variable based on the adopted dataset with its given ranges of system parameter measures. The figure 4 (a, b, c) depicts the shape of the membership functions of the linguistic values of the FS1 parameters. The figure 5 (a, b, c) depicts the shape of the membership functions of the FS2 parameters' linguistic values. The membership functions of the single output of the fuzzy model FS3 are shown in Figure 6.

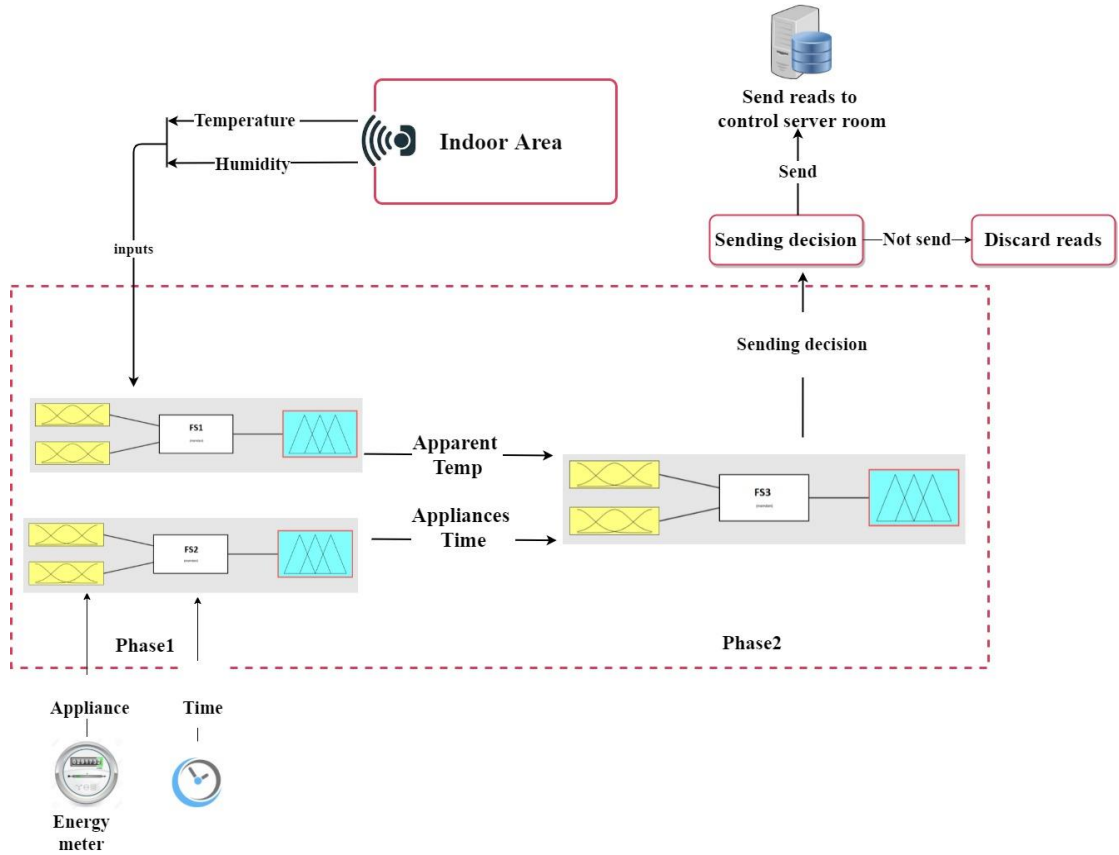

Figure 3. The Implemented fuzzy model approach for temperature and humidity monitoring system

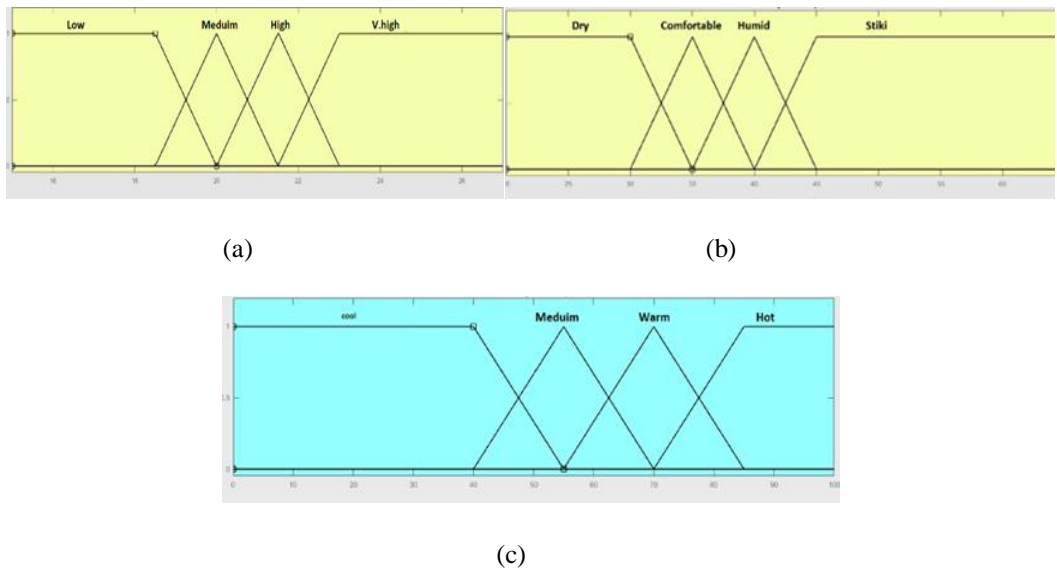

(a)                                    (b)

(c)

Figure 4. Membership functions of the parameters of FS1: (a) for temperature degrees,(b) for humidity degrees, and (c) for apparent temperature degrees

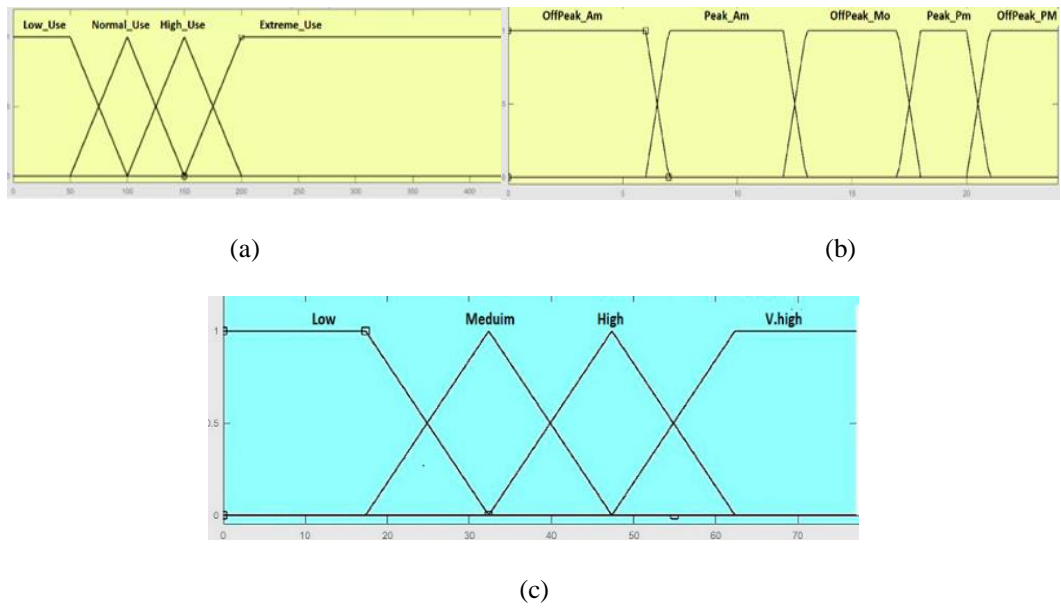

(a)                                                                  (b)

(c)

Figure 5. Membership functions of the parameters of FS2: (a) for appliance energy degrees, (b) for the time of reading, and (c) for appliance time degrees

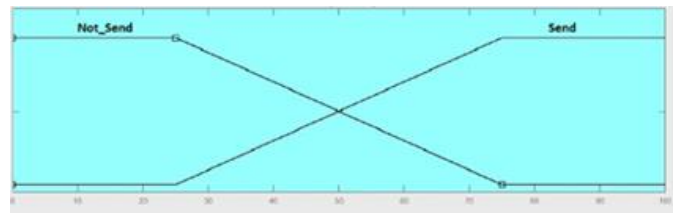

Figure. 6. Membership functions of the output of FS3

All system parameters are summarized in Table 2 with their corresponding forms of fuzzy membership functions.

Table 2. Membership functions: Definition of the fuzzy linguistic values of system parameters.

| System Parameters (Linguistic Variables) | Linguistic Values | Support of Linguistic values |
|---|---|---|
| **Indoor temperature** | low, medium, high, v.high | [0, 20] [18.5, 21.5] [20, 23] [21.5, 100] |
| **Indoor humidity** | dry, comfortable, humid, stiki | [0, 0.35] [0.3, 0.4] [0.35, 0.45] [0.4, 1] |
| **Appliance energy** | low, normal, high, extreme-use | [0, 100] [50, 150] [100, 200] [150,1000] |

| Time of read | OffPeak–AM<br>Peak-AM,<br>OffPeak-Mo,<br>Peak-PM,<br>OffPeak-PM | [12 AM, 7AM]<br>[6 AM, 1 PM]<br>[12 PM, 6 PM]<br>[5 PM, 9 PM]<br>[8 PM, 12 AM] |
|---|---|---|
| Indoor apparent temperature | Cool,<br>medium,<br>warm,<br>hot | [0, 55]<br>[40, 70]<br>[55, 80]<br>[70, 100] |
| Appliance Usage time | Low,<br>medium,<br>high,<br>v.high | [0, 40]<br>[20, 60]<br>[40, 80]<br>[60, 100] |
| Sending decision | Send,<br>not send | [0, 75]<br>[25, 100] |

## 4.4. Operation of the Proposed Fuzzy Model

Based on the proposed groups of linguistic values, we obtain three banks of fuzzy if-then rules for the three fuzzy subsystems. Table 3 shows the bank of 16 rules for the first fuzzy subsystem FS1 of the input phase. The crisp inputs of indoor temperature and humidity are processed by FS1, and the defuzzified crisp output apparent temperature is obtained. The fuzzy if-then rules obtained for FS1 are given in Table 3. To show how the rules in the table can be interpreted, the following are a sample of 3 rules:

- IF the temperature is low and humidity is dry, THEN the apparent temperature is cool.
- IF the temperature is high and humidity is comfortable, THEN the apparent temperature is warm.
- IF the temperature is high and humidity is stiki, THEN the apparent temperature is hot.

Figure 7 depicts an example of an indoor apparent temperature that is cool when the temperature is medium and the humidity is comfortable. Figure 8 depicts the 3D graphical representations of the defuzzifierin terms of indoor temperature, humidity, and output indoor apparent temperature.

Table 3. The obtained bank of fuzzy rules of FS1.

| Temp<br><br>Humid | Low | Medium | High | V.High |
|---|---|---|---|---|
| Dry | Cool | Cool | Medium | Warm |
| Comfortable | Cool | Cool | Warm | Hot |
| Humid | Cool | Medium | Hot | Hot |
| Stiki | Medium | Warm | Hot | Hot |

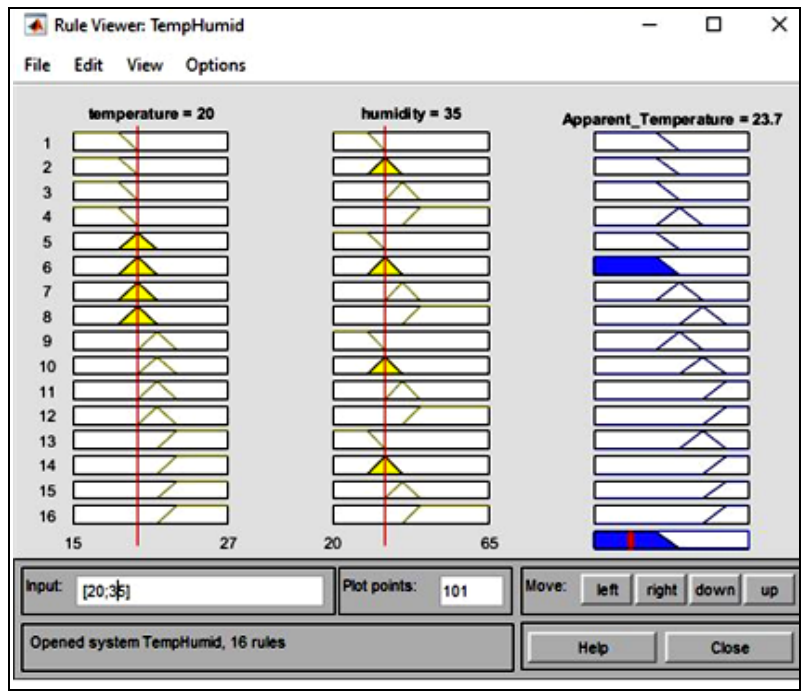

Figure 7. Output: indoor temperature is medium, and indoor humidity is comfortable, then the apparent temperature will be cool

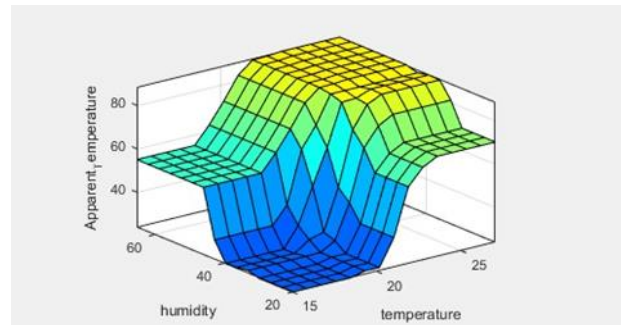

Figure 8. The 3D graphical representations of the Defuzzifier between indoor temperature, indoor humidity, and results of indoor apparent temperature

The second fuzzy subsystem FS2 of the input phase has a bank of 20 rules, as given in Table 4. Through FS2, the crisp inputs appliance energy, and time reads are processed.The defuzzified crisp output appliance usage time is then obtained.The fuzzy if-then rules obtained for FS2 are given in Table 4. To show how the rules in the table can be interpreted, the following are a sample of 3 rules:

- IF the appliance energy is low and the time of reading is off-peak (Am), THEN appliance usage time is low.
- IF the appliance energy is normal and time of reading is a peak (Am), THEN appliance usage time is high.
- IF the appliance energy is high and the time of reading is off-peak (Mo), THEN appliance usage time is medium.

Figure 9 represents an example when the appliance energy is extreme and the time of reading is at Peak (AM) (appliance usage time is v. High). Figure 10 shows the 3D graphical representations of the defuzzifier between appliance energy, time of reading, and output appliance usage time.

Table 4. The obtained bank of fuzzy rules of FS2.

| Appliance energy \ Time | Low | Normal | High | Extreme |
|---|---|---|---|---|
| **OffPeak–Am** | Low | Low | Medium | High |
| **Peak-Am** | Medium | High | V.High | V.High |
| **OffPeak-Mo** | Low | Low | Medium | High |
| **Peak-PM** | Medium | High | V.High | V.High |
| **OffPeak-PM** | Low | Low | Medium | High |

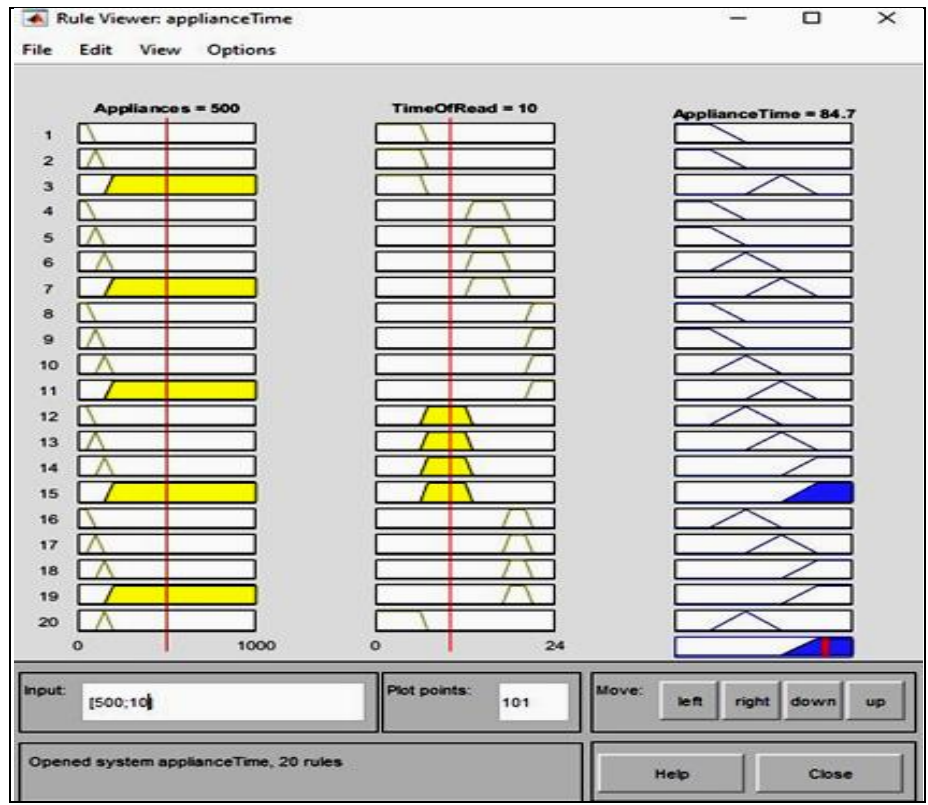

Figure 9. Output: the appliance energy is extreme and the time of reading is Peak (AM), then appliance time will be V.high

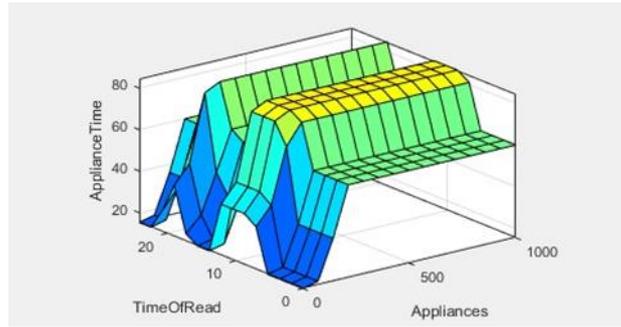

Figure 10. The 3D graphical representations of the Defuzzifier between appliance energy, time of reading, and results of appliance usage time

In phase 2, the output fuzzy subsystem FS3 has 16 rules, as given in Table 5. FS3 receives the crisp outputs of FS2 and FS3, then the defuzzified output sending decision is generated.The fuzzy if-then rules obtained for FS3 are given in Table 5. To show how the rules in the table can be interpreted, the following are a sample of 3 rules:

- IF the apparent temperature is cool and appliance usage time is medium, THEN sending decision is not sent.
- IF the apparent temperature is warm and appliance usage time is high, THEN sending decision is sent.
- IF the apparent temperature is hot and appliance usage time is high, THEN sending decision is sent.

Figure 11 shows an example of a sending decision output when the indoor apparent temperature is medium and the appliance usage time is high. Figure 12 depicts 3D graphical representations of the defuzzifier to indoor apparent temperature, appliance usage time, and output-sending decision.

Table 5. The obtained bank of fuzzy rules of FS3.

| Appliance-usage time / App-Temperature | Low | Medium | High | V.High |
|---|---|---|---|---|
| Cool | Send | Not Send | Not Send | Not Send |
| Medium | Send | Send | Not Send | Not Send |
| Warm | Send | Send | Send | Not Send |
| Hot | Send | Send | Send | Send |

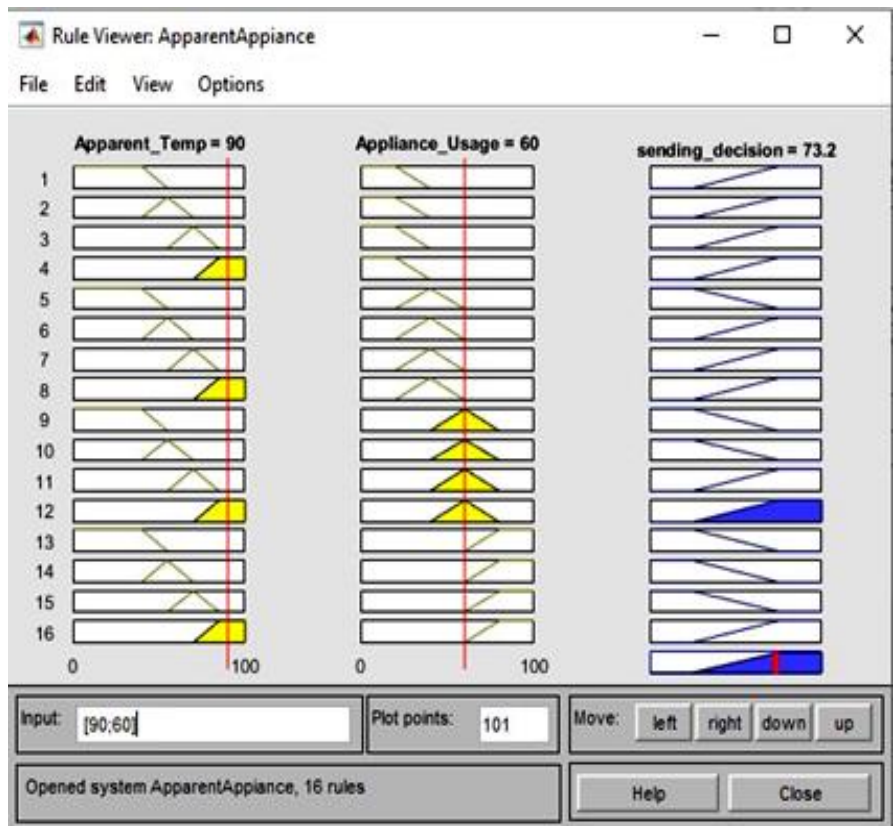

Figure 11. Output: if the indoor apparent temperature is medium and appliance usage time is high, then the sending decision will be sent

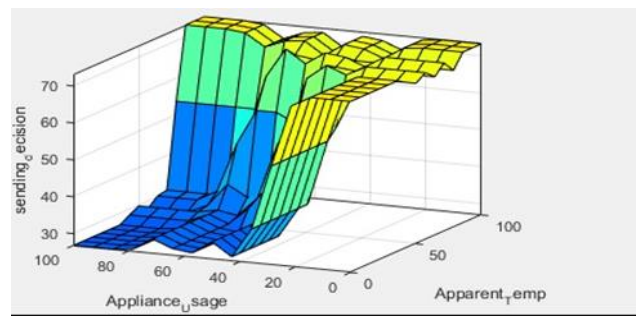

Figure 12. The 3D graphical representations of the Defuzzifier between indoor apparent temperature, appliance usage time, and results of sending decision

## 5. RESULTS AND DISCUSSION

The proposed fuzzy-based monitoring approach is evaluated by comparing its performance with the traditional monitoring approach. The total number of temperature and humidity observation records transmitted between the monitoring system and the server is used to calculate performance. The transmission energy can then be calculated in each case. Figure 13 compares the total energy in joules consumed for total transmissions over the entire period of data recording. The solid column represents traditional energy consumption, which was approximately 957.8 joules, while the total energy consumption using the proposed method appeared in a dashed column and was approximately 844.9 joules. Figure 14, on the other hand, depicts the comparison in a cumulative format,where a solid curve is used for a traditional approach and a dashed curve for the proposed approach.

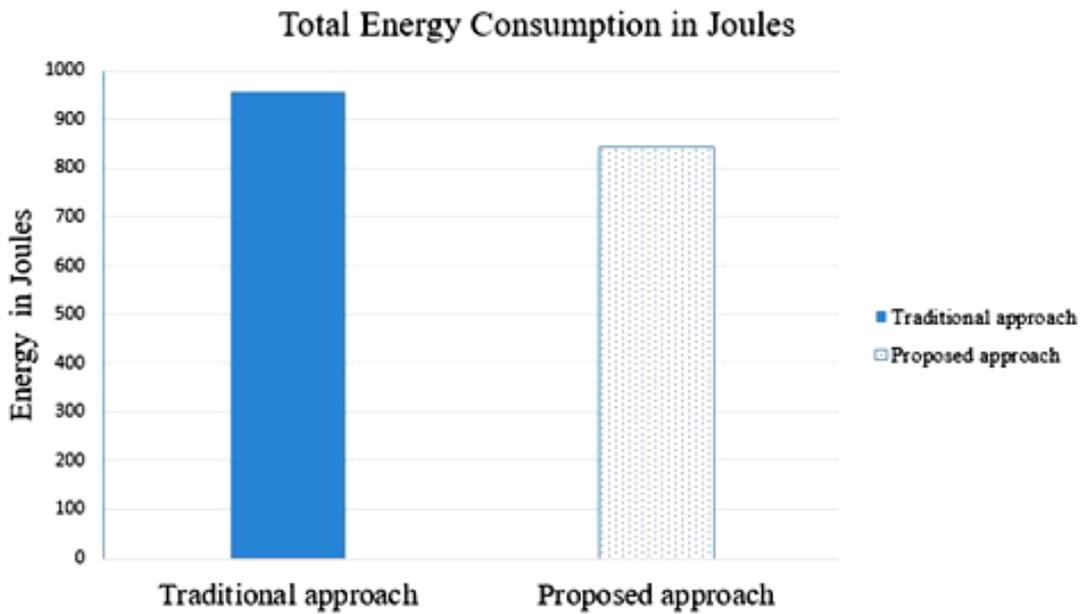

Figure 13. The overall consumed energy in joules

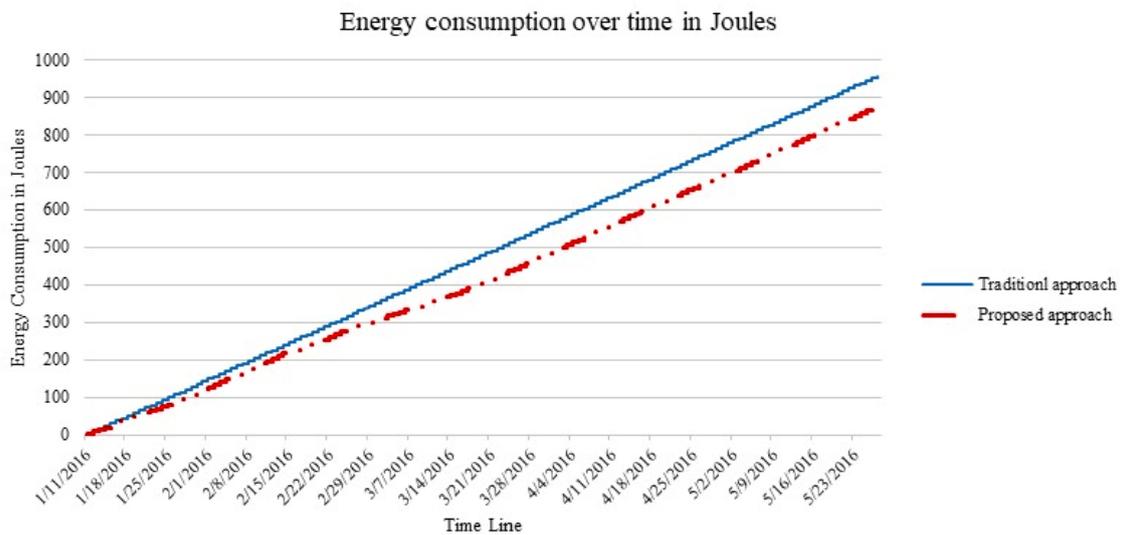

Figure 14. Energy consumption for total transmissions in joules

Table 6 summarizes the performance comparison of the two monitoring approaches. The proposed fuzzy monitoring approach saves a significant amount of transmission energy. When compared with the traditional IoT monitoring approach, the proposed fuzzy model saves 11.8% on transmission energy.

Table 6. The performance of the proposed fuzzy model approach
and traditional approach.

|  | **Traditional Approach** | **Proposed Approach** |
|---|---|---|
| The total number of transmissions | 19735 | 17410 |
| Energy consumption of total transmissions in joules | 957.8 | 844.9 |

By comparing the proposed fuzzy approach with previous approaches of monitoring temperature and humidity that transmit data to external storage, it is clear to find out that the proposed approach is concerned with saving energy consumption unlike a lot of other works as appeared in [24,25]. In addition, the proposed fuzzy approach is based on a customized software solution that can cope with any changes in the monitoring system of temperature and humidity contrary to some other research works, as mentioned in [26,27], which depended on particular low power hardware parts that are not easily found or customized. Also by comparing to solar energy-harvesting technology that is used [28], it is clear that the proposed approach is more reliable and consistent to apply and use.

## 6. CONCLUSIONS

The increasing demands of IoT monitoring systems ask for methods and techniques to reduce energy consumption, which is missed by most of the current remote temperature and humidity monitoring systems. For that purpose, a fuzzy logic approach is proposed in this paper. The adoption of fuzzy logic helps to determine whether the captured data records of temperature and humidity degrees should be transmitted to the control room's server or not according to the state of the surrounding environment as time and the current energy consumption amount without effect the monitoring consistence. For easy understanding and customization based on the monitoring system requirements, the model used clear fuzzy logic variables and sets. In addition, for simplicity, the fuzzy logic approach has been divided into two phases to decompose the number of fuzzy if-then rules of all inputs. This method directly affects the number of data packets exchanged, resulting in less energy consumed during transmission. A simulation experiment using the appliance energy prediction benchmark dataset is used to evaluate the proposed approach. The experimental results showed that the energy consumption has been reduced by 11.8% when using the fuzzy IoT monitoring approach rather than the traditional one.

### CONFLICTS OF INTEREST

The authors declare no conflict of interest.